\documentclass[Afour,sagev,times,doublespace]{sagej}

\usepackage{moreverb,url}
\usepackage{paralist}
\usepackage{tabularx}
\usepackage{enumitem}
\usepackage{wrapfig}
\usepackage{caption}
\usepackage{subfig}
\usepackage[colorlinks,bookmarksopen,bookmarksnumbered,citecolor=red,urlcolor=red]{hyperref}
 
\newcommand{\clrb}{\textcolor{black}} 

\newcommand\BibTeX{{\rmfamily B\kern-.05em \textsc{i\kern-.025em b}\kern-.08em
T\kern-.1667em\lower.7ex\hbox{E}\kern-.125emX}}

\def\volumeyear{2022}

\begin{document}

\runninghead{Li et al.}

\title{Compressing and Interpreting Word Embeddings with Latent Space Regularization and Interactive Semantics Probing}

\author{Haoyu Li\affilnum{1}\affilnum{2}, Junpeng Wang\affilnum{2}, Yan Zheng\affilnum{2}, Liang Wang\affilnum{2}, Wei Zhang\affilnum{2}, Han-Wei Shen\affilnum{1}}

\affiliation{\affilnum{1}GRAVITY research group, Department of Computer Science and Engineering, The Ohio State University, US\\
\affilnum{2}Visa Research, US
}

\corrauth{Haoyu Li, GRAVITY research group, 
Department of Computer Science and Engineering, 
The Ohio State University,
Columbus, US.}
\email{li.8460@osu.edu}

\begin{abstract}
Word embedding, a high-dimensional (HD) numerical representation of words generated by machine learning models, has been used for different natural language processing tasks, e.g., translation between two languages. Recently, there has been an increasing trend of transforming the HD embeddings into a latent space (e.g., via autoencoders) for further tasks, exploiting various merits the latent representations could bring. To preserve the embeddings' quality, these works often map the embeddings into an even higher-dimensional latent space, making the already complicated embeddings even less interpretable and consuming more storage space. In this work, we borrow the idea of $\beta$VAE to regularize the HD latent space. Our regularization implicitly condenses information from the HD latent space into a much lower-dimensional space, thus compressing the embeddings. We also show that each dimension of our regularized latent space is more semantically salient, and validate our assertion by interactively probing the encoding-level of user-proposed semantics in the dimensions. To the end, we design a visual analytics system to monitor the regularization process, explore the HD latent space, and interpret latent dimensions' semantics. We validate the effectiveness of our embedding regularization and interpretation approach through both quantitative and qualitative evaluations.
\end{abstract}

\keywords{High-dimensional Data Visualization, Visual Analytics, Neural Networks, Word embedding}

\maketitle

\section{Introduction}

Embedding, a numerical representation of different data entities generated by machine learning (ML) models, has been widely used in many applications, including natural language processing (NLP), anomaly detection, and recommendation systems. For most cases, the embedding of an entity (e.g., a word in NLP) is a set of numerical values, i.e., a high-dimensional (HD) vector. The embeddings of different entities constitute a HD space, i.e., the \textbf{\textit{embedding space}}. Extensive efforts have been devoted to interpreting this space in the past few years~\cite{liu2017visual, li2018embeddingvis, ji2019visual}. This paper focuses on the analysis of word embeddings.

Researchers have proposed to transfer the embeddings into a \textbf{\textit{latent space}} and use their latent representations to conduct further tasks, utilizing various metrics associated with the transformation.
Mohiuddin et al. \cite{mohiuddin2020lnmap} proposed Latent space Non-linear Mapping (LNMap) to transfer word embeddings of two different languages into a common latent space through two autoencoders (AEs) applied on individual languages' embeddings, and align them in that space for word translations via nearest neighbor search (explained later in \autoref{fig:align}). The non-linearity of the transformation helps to align the embeddings better and improves word translation performance. However, as these works focus entirely on alignment accuracy, they often blindly map the embeddings to an even higher dimensional latent space, without caring much about how semantics are really encoded in that space. As a consequence, the \textit{high dimensionality} and \textit{irregularity} of the latent space bring insurmountable challenges to their interpretation and understanding, going against the paramount need for model transparency~\cite{voigt2017eu}, and limiting their deployment to resource-limited equipment (e.g., mobile devices). Given these problems, our main objectives in this work are to (1) regularize and condense the latent representations of word embeddings and (2) visually interpret the above process to help domain experts better understand and validate the regularization.

The embedding transformation process is our focus, which is handled by an AE traditionally, minimizing the reconstruction loss only. To regularize the latent space, we borrow the architecture of variational auto-encoder (VAE)~\cite{kingma2013auto} to substitute the AE, which adds a regularization term to the loss. To control the regularization strength, $\beta$VAE~\cite{higgins2016beta} is a straightforward extension, which introduces a hyperparameter $\beta$ into the loss of VAE to balance the trade-off between reconstruction accuracy and latent space regularity \clrb{(see \hyperref[sec:betavae]{AE and $\beta$VAE} under the Background section)}. In our experiments of using $\beta$VAE for embedding transformation,  we observed an intriguing phenomenon of \textit{\textbf{dimension deprecation}}, i.e., some latent dimensions of $\beta$VAE get deprecated while the model is converging towards its final state.  
The deprecated latent dimensions lose the capability to encode information, and thus, can be safely removed to compress the embeddings. Meanwhile, using latent representations of the non-deprecated dimensions only, we achieve similar performance on different downstream tasks, indicating almost no information loss over the regularization process (\textit{our first contribution}).
However, related to the dimension deprecation phenomenon, many questions remain unanswered. For example, when does the dimension deprecation start? How to differentiate the deprecated dimensions from useful dimensions? Does each of the remaining non-deprecated dimensions encode more semantics?

The answers to the above questions constitute \textit{our second contribution} on visually interpreting word embeddings and their regularization. Unlike the well-labeled image datasets that used to validate the disentanglement effect of $\beta$VAE~\cite{higgins2016beta, higgins2017scan, wang2020scanviz}, 
the factors that decompose the semantics of word embeddings are often unknown and hard to be extracted. As a result, it is hard to quantify the disentanglement level of latent space. Moreover, ML practitioners often care more about certain semantics in the latent representations when working on domain-specific problems (e.g., gender semantics is of more interest when studying gender bias in word embeddings). To meet the customized semantics exploration needs, we propose to probe the semantics' encoding-level on a latent dimension by interactively perturbing the dimension's value, regressing the reconstructed embeddings (via the decoder of $\beta$VAE), and measuring the angle between the customized semantics and the regressed reconstructions.

Both contributions lead to the need for a visual analytics (VA) system, where users can observe the regularization process of latent space, discern the deprecated and useful latent dimensions, and probe the semantics of individual dimensions. 
Our VA system consists of three major components. The first one provides an overview of the model training dynamics, enabling users to inspect the latent regularization process and catch the point where the regularization starts. The second component adopts ``focus+context'' to present a large number of latent dimensions so that the deprecated and useful dimensions can be easily identified. Also, this component allows users to interactively propose customized semantics (concretized by a pair of words) and reflects the encoding-level of the semantics across latent dimensions through glyphs.
The final component provides necessary details of a user-selected latent dimension for interpretation, e.g., textualizing the semantics along the dimension by dimensionality reductions or reflecting the density of the latent space along the dimension by word clouds.
These tightly coupled components also help to compare the latent space generated from different models (e.g. an AE and a $\beta$VAE) and different languages.

In summary, the contributions our work are as follows:
\begin{compactitem}
\item We initiate the attempt of borrowing $\beta$VAE to compress and regularize the latent representations of word embeddings, and quantitatively validate its effectiveness.
 
\item We design a visual analytics system to explore the HD latent space of word embeddings and probe individual latent dimensions' encoding-level of user-proposed semantics.

\item From case studies with domain experts and comparisons between models, we provide insights into the compression, regularization, and interpretation of latent embeddings.
\end{compactitem}

\section{Related Works}

Our work contributes to the visualization of HD data and ML models. We, therefore, review related works from these two perspectives.

\textbf{\textit{HD Data Visualization.}} Two groups of visualization techniques are widely used to handle the high dimensionality of HD data. The first group directly visualizes all dimensions, e.g., parallel coordinates plot (PCP)~\cite{heinrich2013state} and scatter-plot matrix~\cite{elmqvist2008rolling}. These techniques are intuitive and easily comprehensible but can handle up to $\sim20$ dimensions due to the large consumption of screen space. \clrb{For even higher dimensional data, the second group of techniques, for example, RadVis\cite{hoffman1997dna} and star coordinates\cite{kandogan2000star}, combines dimensionality reduction algorithms (e.g., PCA, tSNE~\cite{van2008visualizing} and UMAP~\cite{mcinnes2018umap}) and scatterplots to reduce the dimensionality and visualize the data in 2D only.} The drawbacks of this group of techniques are the less-interpretable reduced dimensions and the intricate dimensionality reduction process, which may have distorted the data space. In this work, we introduce a zoomable PCP adopting the focus+context exploration strategy to intuitively demonstrate HD embedding data with reasonable space cost. We also employ dimensionality reduction algorithms and scatterplots for the detailed exploration of a locally perturbed region of the HD space. The projected data points in a scatterplot preserve their relative distances, helping to interpret the semantics extensions in the local region and the transformation between a word-pair representing the studied semantics.

\textbf{\textit{Visualization for Machine Learning.}}
There are a plethora of works focusing on using visualization techniques to interpret, diagnose, and/or improve machine learning models~\cite{choo2018visual}. According to a very recent survey~\cite{yuan2021survey}, \clrb{we can roughly categorize them into three groups focusing on (1) improving data quality \textit{before} model building \cite{9308631}, (2) interpreting/diagnosing model dynamics \textit{during} model building \cite{Liu2017},} and (3) analyzing models' outcome \textit{after} model building to evaluate the models or interpret what they have learned~\cite{ren2016squares, liu2017visual}. Our work fits well into the third category of this taxonomy and our focused model outcomes are the HD word embeddings and regularized latent spaces. 
The main objective is to interpret how semantics are encoded in the embeddings or latent spaces. For \textit{word embeddings interpretations}, different visualization solutions have been proposed, solving problems including analogy interpretation~\cite{liu2017visual}, embedding debiasing~\cite{rathore2021verb}, embedding comparison~\cite{li2018embeddingvis}, etc. Moreover, there are visualization works focusing on the concepts defined in the embedding space. They either used human knowledge and user interactions to understand and refine the semantic concepts \cite{el2019semantic} or interactively built concepts to avoid problems by building concepts from seed term set of limited sizes~\cite{park2017conceptvector}.

For the \textit{interpretation of latent spaces}, the majority of visualization works focus on latent spaces from deep learning models with an encoder-decoder architecture, e.g., AE or VAE. For example, SCANViz~\cite{wang2020scanviz} visualized and compared latent space encoded symbolic instructions and another latent space encoded visual concepts to interpret how symbols and concepts are associated in the joint space over training. Latent space cartography~\cite{liu2019latent} defined attribute vectors to denote the semantics offset between a pair of data instances and visually interpreted and verified the pair-wise analogy in latent spaces. 
There are also visualization works that perturb the latent space for model diagnosis and improvements. For example, Wang et al.~\cite{wang2019deepvid} perturbed the latent vector of an instance to be diagnosed to generate its neighbors, and used those neighbors to train an interpretable surrogate model to mimic the original model's local behavior for interpretation.
Gou et al.~\cite{gou2020vatld} perturbed the latent representations to generate data instances with fewer representative features and used them to improve their model's performance.
In this work, we also adopt perturbation-based methods, but to investigate the semantics' encoding-level of different latent dimensions. Also, using our measure of the semantics' encoding-levels, we can compare the latent spaces from different models, and thus, assess the regularization quality of latent space. 

\section{Background}
This section provides ML background for $\beta$VAE and the quantitative metrics for evaluating word embeddings' quality.

\subsection{AE and $\beta$VAE}
\label{sec:betavae}
\begin{figure}
 \centering
 \includegraphics[width=\columnwidth]{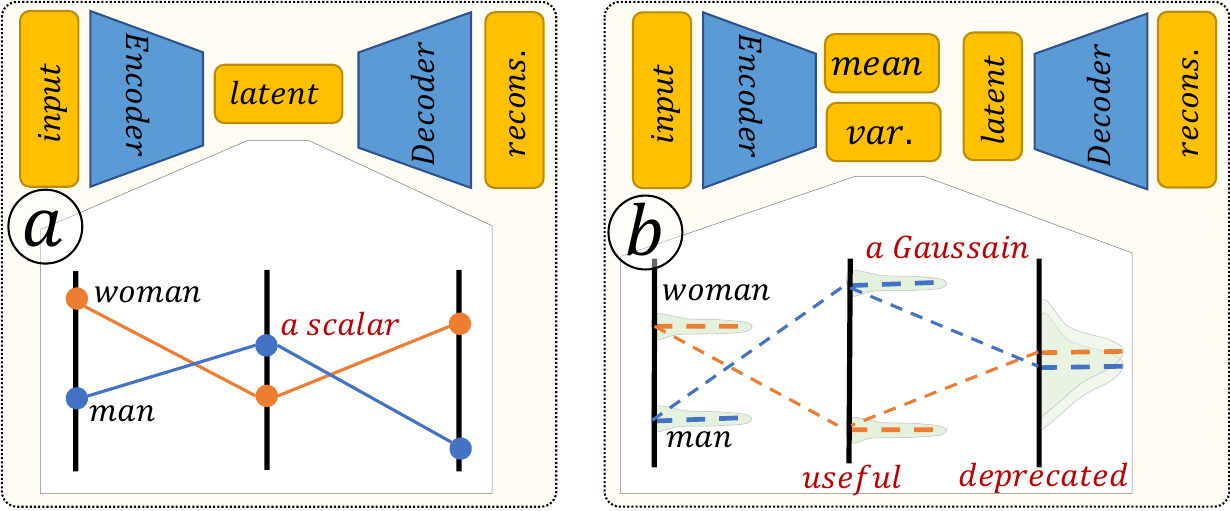}
 \caption{
    (a) AE encodes each instance into a latent vector, i.e., 3 scalars here, since the latent space is in 3D. (b) VAE encodes each instance into a set of Gaussian distributions (each is parameterized by a mean and a variance value). A latent vector is then sampled from them.
 }
 \label{fig:AE_VAE}
\end{figure}

AE (\autoref{fig:AE_VAE}a) is an unsupervised deep neural network constituted of two sub-networks: an encoder and a decoder. The encoder takes each training instance as input and transfers it to an HD latent vector. The decoder takes the vector as input and reconstructs the input instance back. The two sub-networks are trained jointly to minimize the difference between the input and the output of the decoder.

\setlength{\belowcaptionskip}{-15pt}
\setlength{\belowcaptionskip}{0pt}

Similar to AE, VAE~\cite{higgins2016beta} (\autoref{fig:AE_VAE}b) also transfers each input instance into a latent space (the encoder part). However, the latent representation here is no longer a set of scalars (i.e., a vector), but a set of Gaussian distributions, parameterized by a mean and a variance vector. The decoder draws a random sample from each Gaussian to compose the latent vector and uses it to reconstruct the input back. Apart from the \textit{reconstruction loss} inherited from AE, VAE also has a \textit{regularization loss} that minimizes the Kullback–Leibler divergence ($D_{KL}$) between each latent Gaussian distribution and a unit Gaussian distribution. Mathematically,
\begin{equation}
\label{eq:vaeloss}
    \mathcal{L}=\sum_{i=1}^n(x_i-\hat{x_i})^2+ \beta \sum_{i=1}^m D_{KL}\big(\mathcal{N}(\mu_i,\sigma_i^2) \| \mathcal{N}(0,1)\big),
\end{equation}
where $x$ and $\hat{x}$ are the input and reconstructed output with $n$ dimensions and their difference is minimized by \clrb{a $L^2$-norm loss function} (the first term, reconstruction loss). $m$ is the dimensionality of the latent space and the latent representation of $x$ on each latent dimension is a Gaussian parameterized by $\mu_i$ and $\sigma_i^2$. The second term (regularization loss) constrains the latent space by pushing each Gaussian to be a unit Gaussian. 
\clrb{There is a theoretical study \cite{burgess2018understanding}} on how a unit Gaussian constraint can regularize the latent space of VAE  . In short, this constraint term (the second term of \autoref{eq:vaeloss}) can be considered as an information bottleneck (from an information theoretic point of view), which filters out redundant information for the reconstruction, making the latent space more compact and smoother.
$\beta$ controls this bottleneck and balances the two loss terms. It equals 1 in VAE.

A further study on the $\beta$ value derives $\beta$VAE~\cite{higgins2016beta}, where $\beta$ is not necessary to be 1 (and usually larger than 1 to further regularize the latent space). For example, studies\cite{higgins2016beta, burgess2018understanding} have shown that by carefully choosing the $\beta$ value, one can effectively disentangle the latent space to make each latent dimension encode more orthogonal information. Intuitively, this is to further regularize individual latent dimensions, compelling some of them to converge to unit Gaussians. As a result, the remaining dimensions will have to encode more information and in a more effective way to keep the overall loss small.

\subsection{Word Embedding Quality Measurement}
The quality of word embeddings is often evaluated by applying them to different downstream tasks and measuring the tasks' performance. Here we give some examples, covering both monolingual evaluation and cross-lingual evaluation metrics.

\subsubsection{Semantic Similarity and Analogy Scores}
\label{sect:sim_ana}
Monolingual evaluation measures how well certain relations between words are preserved in the embeddings of a single language. This is a supervised evaluation, i.e., we need to have word-pairs with known relations beforehand. \textit{\textbf{Semantic similarity}} and \textit{\textbf{analogy score}} are two widely-used metrics. \textit{SemEval 2017}\cite{camachocollados-EtAl:2017:SemEval} and \textit{Google Analogy Test Set}\cite{mikolov2013word2vec} are the example evaluation sets for the two metrics.

\textit{SemEval 2017} provides the ground truth semantic similarity between 500 word pairs obtained from lexical resources (e.g., WordNet and BabelNet). The evaluation task is to calculate the Spearman correlation coefficient between the ground truth similarity and the similarity from the word embeddings. Therefore, SemEval score should lie in the range of [-1,1], the larger the better.

\textit{Google Analogy Test Set} provides 19,544 sets of coupled word pairs, between which a proportional analogy holds, e.g., Tokyo is to Japan as Paris is to France. We use the corresponding words' embeddings to answer the analogy questions and calculate the correct rate (i.e., accuracy) according to this ground-truth dataset.

\subsubsection{Bilingual Lexicon Induction (Word Translation)}
\label{sect:wt}
Cross-lingual evaluation measures embeddings' quality through tasks that involve embeddings from more than one language. Word translation, or more formally \textit{bilingual lexicon induction} (BLI)~\cite{conneau2017word}, is a good example. The embeddings of two languages are first generated separately and then aligned into a common space. Translating a word from one language to the other can then be conducted by nearest neighbor (NN) search in the space. A translation is considered correct if the right counterpart word appears in the 1NN, 5NN, and 10NN (i.e., measure the accuracy with different levels of tolerance).

Generally, the algorithms for embedding space alignment can be categorized into \textit{supervised/unsupervise}d methods based on whether they use a ground truth dictionary, or \textit{linear\cite{mikolov2013exploiting}/non-linear} methods based on if the mapping function is linear or not. A thorough review of these works can be found from a recent survey~\cite{ruder2019survey}. For our work, we picked one supervised non-linear method, called LNMap~\cite{mohiuddin2020lnmap} for evaluations, because, (1) the work is published very recently and reflects the state-of-the-art performance, (2) the work adopts an AE for the non-linear transformation and generates a higher dimensional latent space, i.e., from 300 dimensional (300D) embeddings to 350D latent space, where we can easily substitute it with our compressed and regularized latent space from a $\beta$VAE for comparison.

\begin{figure}
 \centering
 \includegraphics[width=0.8\columnwidth]{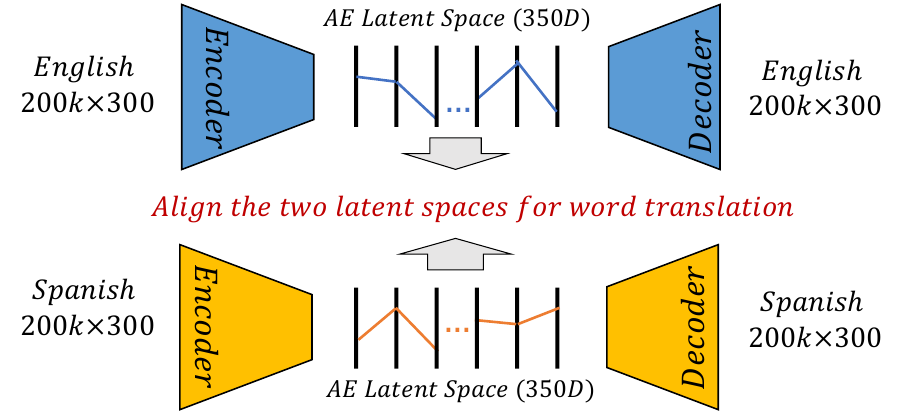}
 \caption{LNMap~\cite{mohiuddin2020lnmap} aligns two languages' embeddings (200k English and Spanish words) by (1) transferring each into a latent space through an AE and (2) aligning the two AEs' latent space. The input and latent space have 300 and 350 dimensions respectively.
 }
 \label{fig:align}
\end{figure}

For two languages' embeddings to be aligned (e.g., English and Spanish in~\autoref{fig:align}), LNMap first trains two AEs, one for each language's embeddings, and then uses the latent representations of the two AEs for alignment (through two non-linear mapping functions and fine-tunings of the encoder of the two AEs).
We will focus on the first step to show how we can compress and regularize the latent representations (i.e., replace the AEs with $\beta$VAEs), without significantly sacrificing the alignment accuracy.

\setlength{\belowcaptionskip}{-15pt}
\setlength{\belowcaptionskip}{0pt}

\section{Embedding Regularization, Semantics Probing}
\label{sec:overview}
This section provides a technical overview of the two main contributions of our work, i.e., compressing and regularizing word embeddings through $\beta$VAE  and probing the semantics of individual latent dimensions via interactive perturbation.

\subsection{Dimension Deprecation Phenomenon}
\label{sect:dim_dep}

LNMap~\cite{mohiuddin2020lnmap} uses 300D FastText embedddings~\cite{bojanowski2017enriching} as the training input, but the latent space of its AE is 350D (higher than the input, \autoref{fig:align}). Although achieving high alignment accuracy, this setting triggered our thoughts that the latent space is not compact, and redundant information may be introduced during the encoding process.
Targeted on regularization, we replaced the AE in LNMap with $\beta$VAE. We experimented with different $\beta$ values and found a small $\beta$ could maintain the quality of the reconstructed embeddings. Also, through some random explorations of $\beta$VAE's latent space, we found the encoded representations of some latent dimensions are very close to unit Gaussians. This is because the regularization loss compels these dimensions to converge to unit Gaussians, dampening their capability to encode any useful information. We therefore call these dimensions \textbf{\textit{deprecated dimensions}}, as they are deprecated by the $\beta$VAE model. 
For example, we use the third latent dimension in \autoref{fig:AE_VAE}b to illustrate a deprecated dimension, where the representations of all different words (e.g., \texttt{man} and \texttt{woman}) are close to a unit Gaussian and they show little difference. In contrast, the other two latent dimensions are useful, where different words are encoded into Gaussians with different means and very small variance (i.e., the narrower shaded region).
When decoding the embeddings back using only the non-deprecated latent dimensions, we still get good reconstruction quality, validating that the deprecated dimensions are indeed superfluous. Moreover, we found the number of deprecated dimensions correlates with the value of $\beta$.

We, as well as our collaborated ML experts, are enlightened by this dimension deprecation phenomenon. However, apart from the random explorations of the latent dimensions, we lack a way to effectively identify all deprecated ones. The ML experts also expressed their requirements of visually disclosing when this phenomenon starts and how it evolves along with the model convergence process.

\subsection{Semantics Probing via Perturbation}
\label{sect:semantic_probing}
Based on the information conservation law, we argue that the remaining useful dimensions (of $\beta$VAE) after removing deprecated dimensions should encode more information compared to the original latent dimensions (of AE). However, there is no readily applicable method to rigorously quantify the amount of information encoded in each latent dimension. Since most of the embedding-related tasks focus on investigating the embeddings' semantics, we propose a perturbation-based method to interactively probe the encoding-level of a given semantic in individual latent dimensions.

\setlength{\belowcaptionskip}{-12pt}
\begin{figure}
 \centering
 \includegraphics[width=\columnwidth]{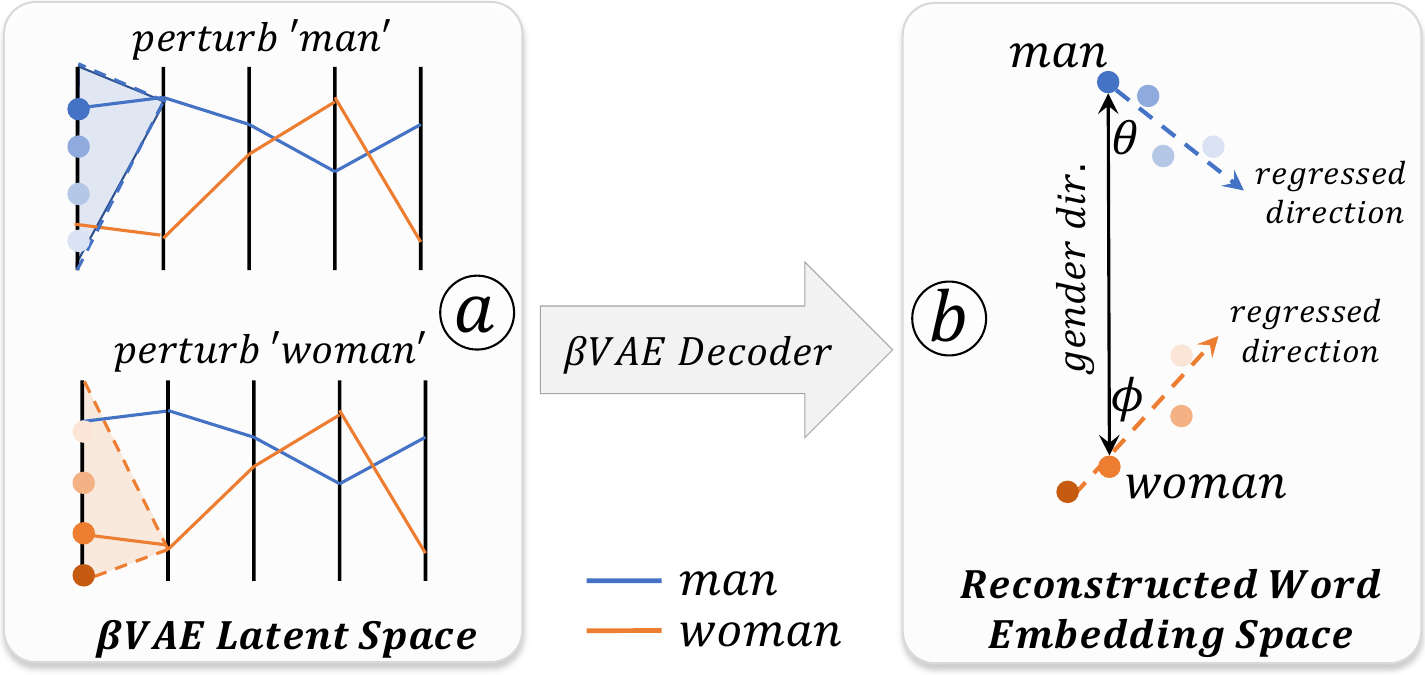}
 \caption{
    (a) Probing the encoding-level of the gender semantics (represented by \texttt{man}-\texttt{woman}) in the first latent dimension. (b) $(\theta{+}\phi){/}2$ denotes the encoding-level of gender semantics in the dimension. 
 }
 \label{fig:probing}
\end{figure}
\setlength{\belowcaptionskip}{0pt}

Semantics is believed to be expressed along a certain direction in the HD latent space~\cite{liu2019latent}, which can be approximated through a representative word-pair. For example, the gender direction can be identified by connecting the point (latent vector) of \texttt{man} and \texttt{woman} in the HD latent space.
In practice, however, as each latent dimension usually mixes multiple semantics, it is nontrivial to reveal which latent dimension is more responsible for what semantics. To address this, our perturbation-based method probes the encoding-level of a user-specified semantics in the following steps (\autoref{fig:probing}):

\begin{compactitem}

\item For a focused latent dimension and a pair of words (\texttt{man}-\texttt{woman}), we perturb the values of one word's latent vectors on that dimension, while keeping the other dimensions' value unchanged. For $\beta$VAE, we perturb the mean vector and just use the perturbed mean vector for decoding, as it is the most representative sample. Fig.~\ref{fig:probing}a (top) perturbs the first dimension of the latent vector of \texttt{man}.

\item Feeding the set of perturbed latent vectors (blue points in \autoref{fig:probing}a (top)) into the decoder of $\beta$VAE, we reconstruct a set of perturbed word embeddings (the blue points in \autoref{fig:probing}b). 

\item For the set of reconstructed perturbations (i.e., perturbed embeddings), a representative direction is needed to abstract the semantics encoded along the perturbed latent dimension. We use the first principle component (FPC) from PCA (of the reconstructed perturbations) as a linear approximation for the semantics' direction (i.e., a regressed direction), and the explained variance of the FPC is the semantics' extension.

\item The angle between the semantics' direction represented by the word pair and the regressed direction (from FPC), i.e., $\theta$ in ~\autoref{fig:probing}b, reflects the encoding-level of the first latent dimension \textit{localized} at the word \texttt{man}. This angle should be smaller if the latent dimension encodes more of the probed semantics. Conversely, it will be 90\textdegree{} if the dimension encodes no information of the probed semantics.
Repeating the process to another word, we get another angle, i.e., $\phi$ in ~\autoref{fig:probing}b.

\item Finally, we use $({\theta}+{\phi})/2$ to denote the encoding-level of the user-proposed semantics in the focused latent dimension. 
    
\end{compactitem}

For multiple latent dimensions, we can repeat the above steps to derive their encoding-levels of user-specified semantics. However, it is hard to provide an overview of all dimensions' behaviors through these numerical values only.  
Also, ML practitioners will have different semantics to probe over their exploration with the latent space. Third, we also want to illustrate the semantics encoding discrepancy between useful and deprecated dimensions.
These issues bring the need for an interactive visual analytics system.

\section{Design Requirement Analysis}
\label{sect:requirements}
\clrb{In the process of designing the visual analytics system}, we worked with 4 NLP experts
, an NLP professor, one of his senior Ph.D. students, and two industrial lab researchers. All have multiple years of experience in the generation, disentanglement, compression, regularization, and alignment of word embeddings, and two of them have co-authored this paper. Over the collaborations and the explorations of related experimental results, the need for an integrated VA system quickly arises \clrb{(as we have partially mentioned in the \hyperref[sec:overview]{Embedding Regularization, Semantics Probing} section). }

The design and development of the VA system have gone through three major stages: (1) collecting the looming needs for HD embedding interpretations; 
(2) prototyping a dimension-wise semantics probing and visualization system; (3) iterative design, development, and modification of the system based on experts' feedback. Over this iterative process and the interactions with the NLP experts, we distilled the following requirements form four major perspectives: \textit{\textbf{R1: Model Evolution}}, \textbf{\textit{R2: Latent Space Exploration}}, \textbf{\textit{R3: Semantics Analysis}}, and \textbf{\textit{R4: Model Comparison}}, detailed as follows:\\
\textbf{R1: Provide \textit{model evolution analysis} over the embedding regularization process.} The training of $\beta$VAE involves two contradictory loss terms, quantifying the reconstruction quality and regularization scale. Therefore, it is critical to disclose the dynamic balance between these two parts. In detail, the VA system needs to:
\begin{itemize}[leftmargin=*,noitemsep,topsep=0pt]
    \item \textit{R1.1 Comprehensively reflect the reconstructed embeddings' quality} through different metrics apart from the regularization loss. 
    \item \textit{R1.2 Clearly depict the regularization process}, e.g., when the regularization starts and which dimension becomes deprecated.
\end{itemize}
\textbf{R2: Allow convenient \textit{latent space exploration} and identify interesting latent dimension.} 
Being able to flexibly explore the HD latent space is always a need from the NLP experts, especially when working with $\beta$VAE. During explorations, dimension-wise statistics are useful to guide users towards interesting dimensions, and further probe their semantics. Therefore, our system needs to:

\begin{itemize}[leftmargin=*,noitemsep,topsep=0pt]
    \item \textit{R2.1 Provide overall dimension-wise statistics.} The NLP experts are interested in dimension-wise statistics, e.g., the distribution of mean values (of the encoded Gaussians), as they are informative indicators for useful and deprecated dimensions.
    \item \textit{R2.2 Support easy dimension exploration/selection.} As the dimensionality of the latent space is often high, a quick and easy dimension selection, supported with dimension ordering/filtering, is a must to improve the exploration efficiency.
    \item \textit{R2.3 Enable interactive semantics probing.} The experts usually have an accumulated set of word-pairs for different semantics. Interactive probing and switching between them are needed. 
\end{itemize}
\textbf{R3: \textit{Semantics analysis} for the dimensions of interest.} Focusing on a single dimension of interest, our system needs to provide details for its encoding-level of the studied semantics and the semantics density in different value ranges to help in-depth understandings.

\noindent
\textbf{R4: Provide \textit{model comparison} between AE and $\beta$VAE, and across languages.} This is to comparatively study the advantages of $\beta$VAE over AE, as well as $\beta$VAE's regularization power across different languages' embeddings.
The comparison should cover:

\begin{itemize}[leftmargin=*,noitemsep,topsep=0pt]
    \item \textit{R4.1 Model evolution comparison.} $\beta$VAE is trained with an extra regularization term. It is of interest to understand how this term influences the training compared to an AE.
    \item \textit{R4.2 Latent space comparison.} 
    This comparison focuses on the latent space statistics and semantics of AE and $\beta$VAE, targeting to highlight the advantage of $\beta$VAE.
    \item \textit{R4.3 Comparison across languages.} Our regularization generates different numbers of useful dimensions in different languages. 
    This comparison targets to provide insights into the regularization effects on languages with different richness-levels. 
\end{itemize}

\section{Visual Analytics System}
With the requirements, we developed a VA system with four visualization views, detailed in four sub-sections.

\setlength{\belowcaptionskip}{-12pt}
\begin{figure*}[tbh]
 \centering
 \includegraphics[width=\textwidth]{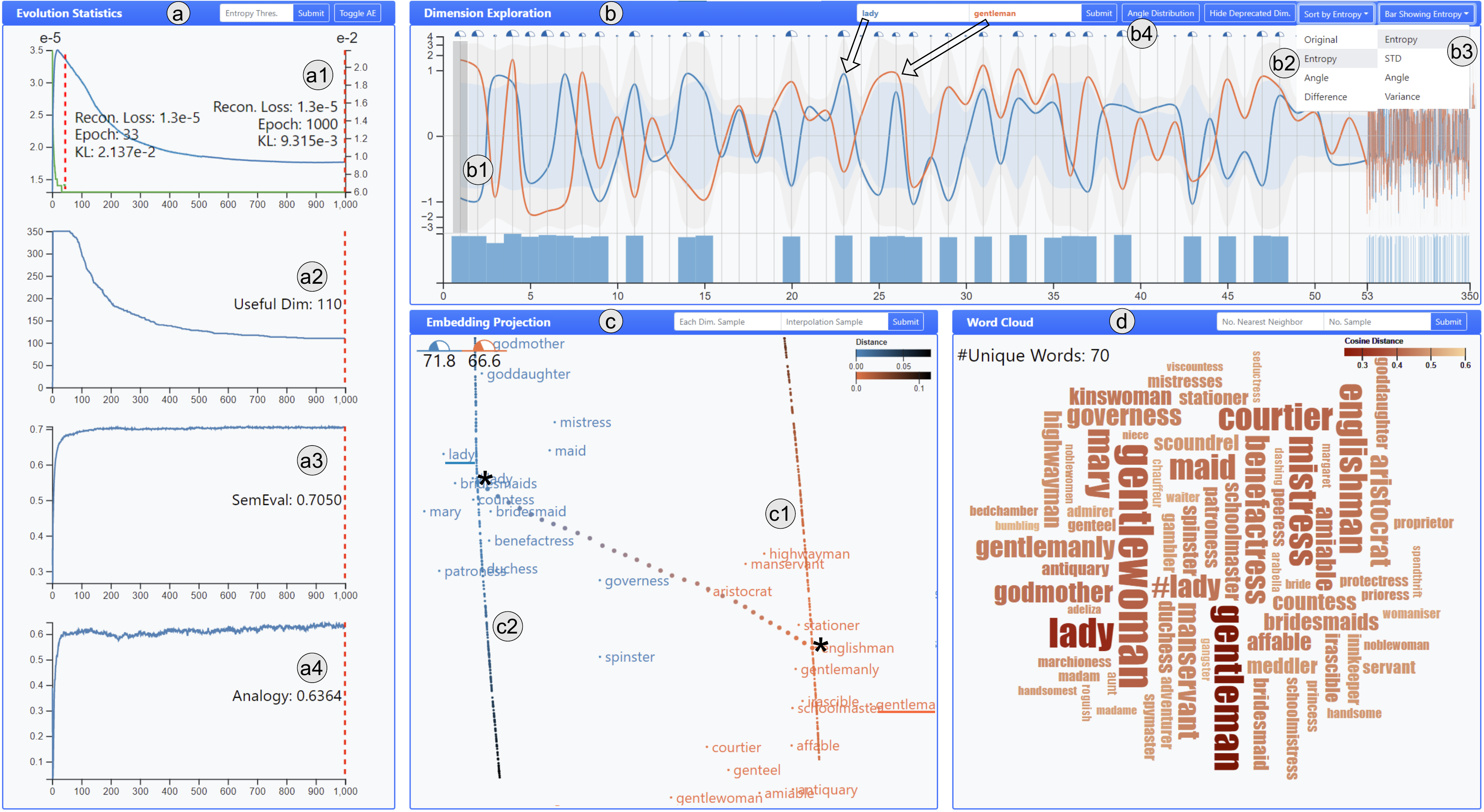}
 \caption{
  Our system consists of four views, (a) the \textit{Model Evolution View} presents five training dynamics reflecting embeddings' quality and regularization scale, (b) the \textit{Dimension Exploration View} employs a zoomable PCP and customized glyphs to present latent dimensions, allowing users to probe the encoding-level of individual dimensions, (c) the \textit{Projection}, and (d) \textit{Word Cloud View} disclose details of the selected dimension (e.g., semantics extension, latent space density) and relate it to the words' semantics.
 }
 \label{fig:system}
\end{figure*}
\setlength{\belowcaptionskip}{0pt}

\subsection{Model Evolution View}
The \textit{Model Evolution View} shows an overview of the training process (\autoref{fig:system}a). The model statistics in this view include the \clrb{reconstruction loss (a1), number of useful dimensions (a2), semantic similarity (a3), and analogy scores (a4)}{, reflecting the latent vectors' quality (R1.1) The statistics also include the KL loss and the number of useful dimensions, reflecting the regularization scale of the latent space (R1.2).
All five statistics are calculated on-the-fly after each epoch and presented in four synchronized line charts. 

Whether a dimension is useful or deprecated is determined as follows. First, we use the $\beta$VAE encoder to encode all $n$ words into the latent space. For each latent dimension, we have $n$ Gaussians (i.e., $n$ pairs of mean and variance), one for each word. We then compute the entropy of the $n$ mean values. For a useful dimension, the corresponding entropy is larger as the $n$ words are encoded using the full range of values on the dimension. For deprecated dimensions, however, their entropy is small and tends to be 0, restricted by the regularization loss (i.e., the KL loss of $\beta$VAE pushes all Gaussians to be unit Gaussians). \autoref{fig:AE_VAE}b shows typical useful and deprecated latent dimensions. We will verify the efficacy of this method later by interactively perturbing the latent dimensions.

\textit{\textbf{Interactions.}} Users can switch between different models, i.e., $\beta$VAE or AE via the ``Toggle AE'' button (\autoref{fig:system}a), and the corresponding statistics will be updated accordingly.
They can also choose a model with the desired number of epochs (the red dashed line marks this selection). The flexible switch between models/epochs facilitates the comparison between them (R4.1, R4.2). 
The selection in this view will be automatically propagated to other views, where users can then explore the latent dimensions and encoded semantics.

\subsection{Dimension Exploration View} 
The \textit{Dimension Exploration View} is designed to explore the HD latent space and probe the semantics encoded in different dimensions (R2). It consists of a zoomable PCP and customized glyphs.

One critical issue to visualize and interact with HD data is to intuitively present all data dimensions and flexibly explore different subsets of dimensions.
We adopt a PCP~\cite{inselberg1990parallel} to solve this issue as shown in \autoref{fig:system}b, where each axis represents one dimension of our HD latent space. To alleviate the visual clutter for a large number of latent dimensions, we make the PCP zoomable. Users can focus on several dimensions by brushing the axes horizontally. Unfocused dimensions will be zoomed out and shown on the sides with less space, e.g., \autoref{fig:system}b focuses on dimensions $1{\sim}53$, and the rest are zoomed out. Our zoomable PCP encodes two types of information. 

The first type presents the value distribution across all words' embedding on each dimension of the latent space (R2.1).
For AE, the encoded representation on each latent dimension is a scalar and we can easily compute its value distribution. However, for $\beta$VAE, the latent representation is a Gaussian. As the mean of a Gaussian is the most representative sample, we use it as the latent representation. For both cases, we compute the value range and confident interval (CI, $1^{st}{\sim}3^{rd}$ quantile) for the set of scalars, and use the gray and blue band in Fig.~\ref{fig:system}b to connect them across axes. Since the latent values mostly fall into [-1, 1], we use an exponential scale, i.e.,  $y=x^{0.3}$, instead of a linear scale, on the vertical direction to enlarge the space between -1 to 1. Additionally, we add a bar chart below the PCP. Each bar is aligned with one axis, and its height encodes the entropy of the corresponding latent dimension's scalar values, which is an indicator of if the dimension is deprecated or not.

The second type of information is related to the user-specified word-pair (R2.2), representing the semantics to be probed with. The two text boxes on the header of Fig.~\ref{fig:system}b allow users to provide a pair of words and the zoomable PCP will show their latent representation as two curves (in blue and orange). Meanwhile, our system will compute the encoding-level of the semantics across all latent dimensions, i.e., compute the angles introduced before, and present them through \textit{semantic angle glyphs} above the corresponding axes. 

The design of the \textit{semantic angle glyphs} is illustrated in Fig.~\ref{fig:variance_example}a. The horizontal direction is the direction of the proposed semantics. The half disk with a filled sector above the horizontal line indicates the encoding-level, i.e., the average of $\theta$ and $\phi$ in \autoref{fig:probing}. The smaller the angle, the higher the encoding-level of the probed semantics.

While exploring the latent dimensions using these glyphs \clrb{(e.g. sorting dimensions by their angle using the widget at \autoref{fig:system}-b2)}, we found some deprecated dimensions have an even smaller angle than useful dimensions, contradicting our belief in those dimensions' behaviors. With a thorough analysis, we realize that we should measure not only the semantics angles but also the regression quality. As shown in Fig.~\ref{fig:variance_example}b, the corresponding latent dimension has a smaller angle (compared to that in Fig.~\ref{fig:variance_example}a). However, most of the reconstructed points are cluttered in a small extent and the regressed direction is not as certain as that in Fig.~\ref{fig:variance_example}a. This is a common phenomenon for deprecated dimensions, and the regressed direction is very susceptible to noisy samples. To reflect this, we use the radius of the half disk to encode the extent of the reconstructions along the regressed direction (i.e., the explained variance of the first principle component in PCA). Latent dimensions with a small extent are usually deprecated, and of less interest.

Our system can also present the angle distribution over dimensions through a histogram. 
We normalize this histogram to make the angle distribution a probability density function (PDF) so that the PDFs for different latent spaces (with different numbers of dimensions) can be compared as two overlaid area-plots. An example is shown in \autoref{fig:angle_distribution}, and the view is shown on-demand when users click the button of ``Angle Distribution'' in \autoref{fig:system}-b4.

\setlength{\belowcaptionskip}{-10pt}
\begin{figure}[tb]
 \centering
 \includegraphics[width=\columnwidth]{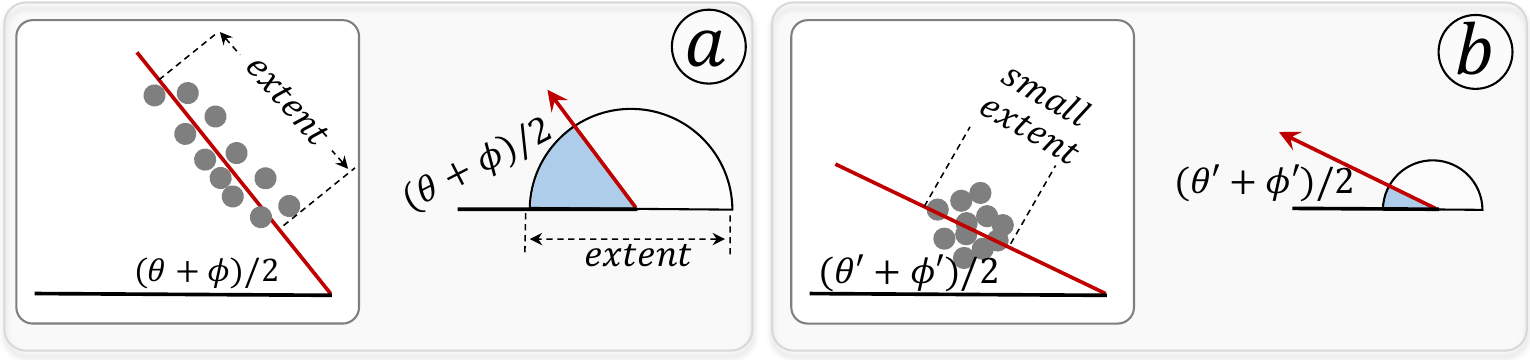}
 \caption{
 A glyph reflects the angle between the directions of the semantics (black) and regressed direction (red). The glyph's radius encodes the extent of the samples along the regressed direction. (a) and (b) represent useful and deprecated dimensions, respectively. 
 }
 \label{fig:variance_example}
\end{figure}
\setlength{\belowcaptionskip}{0pt}

\textbf{\textit{Interactions.}} 
To easily identify interesting latent dimensions (R2.2), \clrb{through the widget at \autoref{fig:system}-b2}, our system allows users to sort the dimensions by (1) their values' entropy, (2) semantics angle, or (3) the latent values' difference between the specified pair of words on each dimension. The dimensions can also be filtered or hidden (e.g., the deprecated ones) from the PCP. 
The zooming capability of the PCP helps users to focus on a few interesting latent dimensions only, from which they can select a dimension or brush a value range on that dimension for further analysis in other views.

\subsection{Embedding Projection View} 
This component (\autoref{fig:system}c) provides a local view of the word embedding space around the specified semantics, helping users to understand the semantics probing process. It uses a scatterplot to present (1) the user-specified pair of words, (2) the two words' neighbors, and (3) the semantic extension of the two words along a latent dimension (selected from the \textit{Dimension Exploration View}).

First, for the user-specified pair of words, we transfer them into the latent space (via the encoder), and generate samples by linearly interpolating the two encoded latent vectors. Lastly, we reconstruct the two words and the interpolated samples back into the embedding space (via the decoder) and use PCA to project them into 2D for visualization. In \autoref{fig:system}c, the reconstructions of the two words \texttt{lady} and \texttt{gentleman} are denoted by the two asterisks. 
The samples between them are colored with interpolated colors, from which, users can perceive the semantic direction specified by the word-pair.

Second, for each of the two reconstructed words, we find its $k$ nearest neighbors in the embedding space (based on cosine-distance), use PCA to project them onto the scatterplot, and label them with the corresponding texts. Showing these neighbors is to textualize the context of the semantics. These neighbors are colored in blue or orange, indicating which word they are the neighbors for. It can be seen that there is an offset between the right asterisk and the original embedding for \texttt{gentleman}, which denotes the reconstruction error.

Third, we also perturb the user-selected latent dimension around the two words, decode the perturbations back to the embedding space, and use PCA to project them to the scatterplot to get the \textit{semantics extensions} around each word (\autoref{fig:system} c1, c2). This visualization intuitively reflects the probing process explained in \autoref{fig:probing}.

While the users of our system are well aware of the dimension distortions introduced by dimensionality reductions, they still appreciate the intuition this view could bring to the understanding of individual latent dimensions. In most cases, we can see linear trajectories formed by the interpolated samples between the two words and the perturbed samples around each of the two words. To accurately reflect the two angles ($\theta$ and $\phi$ in \autoref{fig:probing}b) in the scatterplot, we visualize two angular glyphs on the top-left corner of \autoref{fig:system}c and label the two angles. The color of the glyph matches the color of the perturbed word. We choose PCA for the projection, as we are more interested in the global structure of the three types of data instead of their local structures. However, we have also explored other dimensionality reduction algorithms (see our Appendix).

\subsection{Word Cloud View} 
This component allows users to brush a value range on the selected latent dimension (\autoref{fig:system}-b1), and textualize the semantics in that range through a word cloud (\autoref{fig:system}d), to better understand the HD manifold of the latent space along the selected dimension (R3).

\setlength{\belowcaptionskip}{-15pt}
\begin{figure}[tb]
 \centering
 \includegraphics[width=\columnwidth]{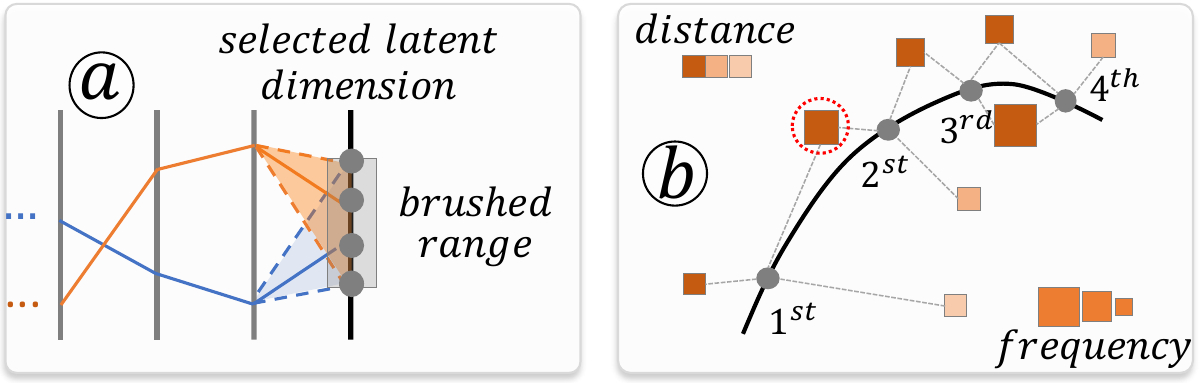}
 \caption{
(a) Perturb a word-pair on in a specified range of the selected dimension. (b) Decode the perturbations into the embedding space, find their $k$NNs, and visualize them through a world cloud.
 }
 \label{fig:cloud}
\end{figure}
\setlength{\belowcaptionskip}{0pt}

The generation of the word cloud is explained in \autoref{fig:cloud}.
From \autoref{fig:cloud}a (the PCP of the \textit{Dimension Exploration View}), $n$ random samples ($n{=}4$) are drawn within the specified range by perturbing the two words, and the samples are decoded into the embedding space, where we find $k$ nearest neighbors for each sample ($k{=}3$ in \autoref{fig:cloud}b). Many neighbors (the rectangles) are shared by the samples (the points), so the total number of neighbors is less than $nk$. 
For each unique neighbor (word), we find its distances to all $n$ samples and use the minimum one to color it.
The smaller the distance, the darker the color. This is to reveal how representative the neighbors can denote the samples' semantics. Meanwhile, we use size to encode the frequency of each word, which is computed as the sum of its inversed rank if it appears in the $k$ nearest neighbors of a sample. For example, for the highlighted square node in \autoref{fig:cloud}b, it is the second nearest neighbor to the $1^{st}$ sample and $k{=}3$ in this example, so its inversed rank is $3{-}2{=}1$. 
The highlighted node is also the first neighbor to the $2^{nd}$ samples. So, its inversed rank is $3{-}1{=}2$. This square is not the top 3 neighbors of the 3rd and 4th samples. So, the total frequency for the highlighted square is $1{+}2{=}3$. More frequent neighbors are presented with larger font size in \autoref{fig:system}d, which helps to denote the dominant semantics in the brushed range.

The number of unique words (i.e., the diversity of the word cloud) reflects the richness of the brushed range, i.e., the gradient of the HD space along the selected latent dimension. The bulk of words also reflect the encoded semantics in the brushed range of the dimension. Using this view, users can easily compare the semantics encoded in different value ranges of the same or different latent dimensions from the same or different models.

\section{Case Study}
\label{sect:case_study}
We run case studies with the four domain experts using our system to interpret the embedding regularization process. After the studies, we conduct open-ended interviews with them and summarize their feedback at the end of this section.

We focus on \texttt{English} embeddings and organize the cases into three topics, model evaluation (R1), latent exploration (R2), and semantics interpretation (R3). For each topic, we first explain the findings from $\beta$VAE and then compare them with the AE counterpart (R4). Both AE and $\beta$VAE are trained using the specifications from the LNMap study\cite{mohiuddin2020lnmap} (i.e., transferring 300D embeddings into a 350D latent space). We set $\beta{=}10^{-5}$ in $\beta$VAE and explain this choice in the evaluation.

\subsection{Model Evolution Analysis}
The first and foremost step when analyzing the latent space of $\beta$VAE is to overview the training process and identify when the model converges and if dimension deprecation happens (R1). These answers can be found from the \textit{Evolution Statistics View} (\autoref{fig:system}a).

For model convergence of the $\beta$VAE (\autoref{fig:system}-a1), we found the reconstruction loss curve (in green) drops and converges after $33$ epochs, whereas the regularization loss curve (in blue) increases in the first few epochs and starts to drop after that.
Our discussions with the domain experts on this observation lead to the proposition that the reconstruction and regularization happened sequentially. We also speculate the small $\beta$ value, i.e., a larger weight on the reconstruction term, contributed to the optimization order. Specifically, guided by the ratio between the two loss terms, the model first converges to an ``optima'' similar to the AE (in the first 33 epochs) to achieve on-par performance. After that, it gradually reorganizes/transfers information from some dimensions to the rest to regularize the latent space (reflected by the drop of the blue curve), while not moving the model out of the ``optima'' (reflected by the flatten green curve).

The statistics shown in other plots echo our observation of the loss curves.
The semantic similarity and analogy score (\autoref{fig:system}-a3, \ref{fig:system}-a4) increase along with the decrease of the reconstruction loss, and converge around 33 epochs, reflecting the good reconstruction quality of the embeddings. The number of useful dimensions (\autoref{fig:system}-a2) stays at 350 and starts to drop after around 33 epochs, reflecting the occurrence of the dimension deprecation phenomena.

Switching to the AE for comparison (R4.1), we found it converges even faster (around 12 epochs, \autoref{fig:AE_statistics}), probably due to the free of restraint from the regularization loss. The final reconstruction loss, semantics similarity (SemEval), and analogy scores are similar to those of the $\beta$VAE, verifying our hypotheses that the $\beta$VAE and AE converge to similar optima that yields good reconstruction quality. However, the number of deprecated dimensions never drops for the AE model (not shown here as the figure is a flattened curve and trivial).

\setlength{\belowcaptionskip}{-15pt}
\begin{figure}[tb]
 \centering
 \includegraphics[width=\columnwidth]{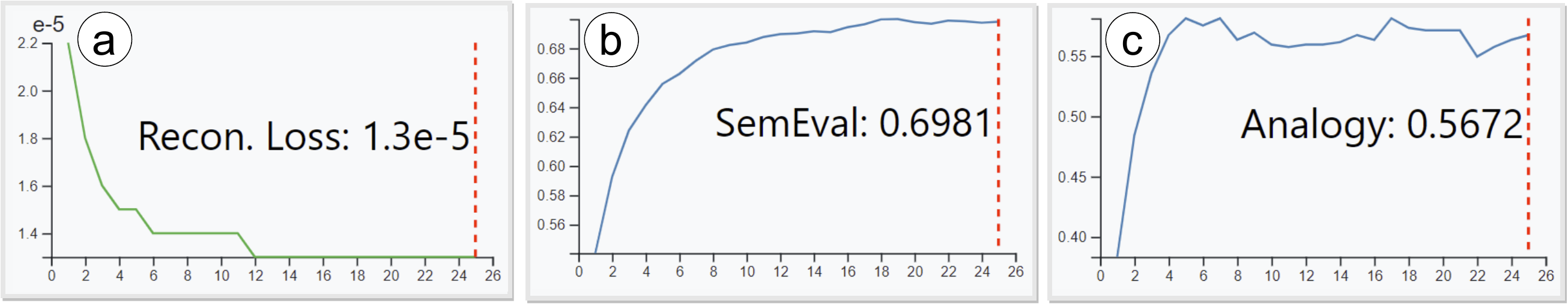}
 \caption{Reconstruction loss, semantic similarity (SemEval), and analogy scores for the AE, which converges in 12 epochs.}
 \label{fig:AE_statistics}
\end{figure}
\setlength{\belowcaptionskip}{0pt}

\subsection{Latent Dimension Exploration}
After an overview of the regularization process, we then select a converged model state to explore the HD latent space (i.e., epoch 1000 for $\beta$VAE and epoch 25 for AE).

\paragraph{\textbf{Dimension Deprecation.}} The first thing we wanted to find was the deprecated dimensions (R2.1), which can be identified by simply sorting the latent dimensions using the entropy of all encoded latent mean values (\autoref{fig:system}-b2). \autoref{fig:empty_latent}a shows the zoomable PCP after dimension sorting. The dark blue bars (\autoref{fig:empty_latent}b) beneath the PCP reflect the entropy of each dimension (in decreasing order). A clear separation of the latent dimensions is observed. The first 110 dimensions have a large entropy and their latent value (i.e., the mean of the encoded Gaussian for different words) ranges from -4 to 4. The rest 240 dimensions have a small entropy (nearly zero) and their value ranges from -0.4 to 0.4. The entropy reflects the information volume. So, we can easily identify the first 110 dimensions as useful dimensions and the rest as deprecated ones. \clrb{Detailed explanation of deprecated dimensions can be found in the section \hyperref[sect:dim_dep]{Dimension Deprecation Phenomenon}}.

Through the widget at \autoref{fig:system}-b3, we can also use the bar chart to reflect the average of all standard deviation values (of the encoded Gaussins on each latent dimension). \autoref{fig:empty_latent}c shows this. It can be seen that useful dimensions (the first 110) encode words into Gaussians of different mean values and small standard deviations, i.e., different words are encoded into \textit{narrower} Gaussians, such that there is no overlap between them (check our schematic illustration in \autoref{fig:AE_VAE}b). In contrast, the deprecated dimensions encode words into Gaussians of similar mean values (around 0) and large standard deviations (around 1), i.e., overlapped unit Gaussians (see \autoref{fig:AE_VAE}b).

Moreover, the differentiation of useful and deprecated dimensions is nearly binary and there are no dimensions in between (with a moderate entropy). Since deprecated dimensions encode little information (a unit Gaussian for all words), we hide them in later explorations. 

\setlength{\belowcaptionskip}{-15pt}
\begin{figure}[tb]
 \centering
 \includegraphics[width=\columnwidth]{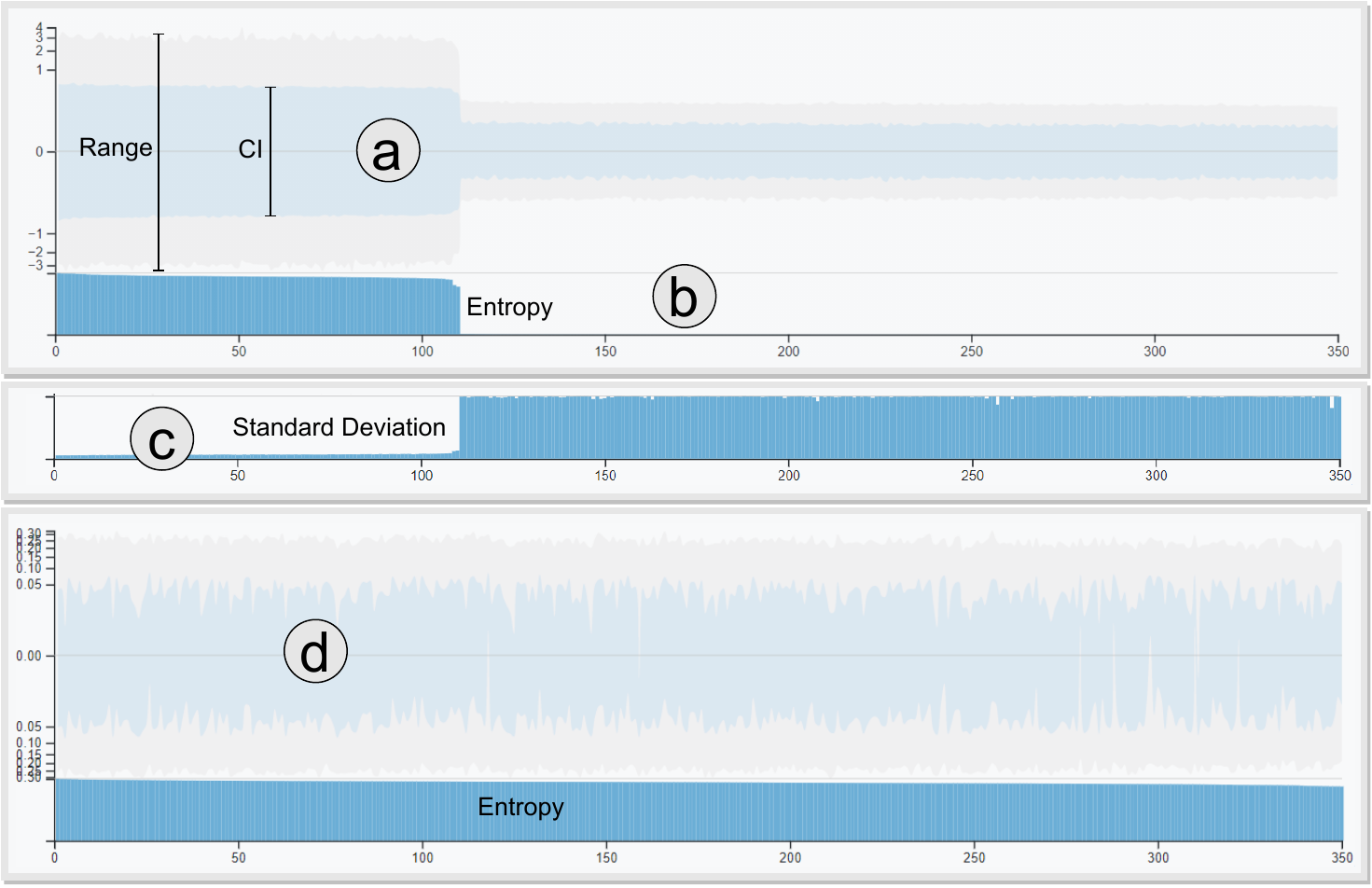}
 \caption{Sorting the latent dimensions in the zoomable PCP by the entropy of mean values. (a) Value range and CI of each dimension; (b, c) Bar charts encoding the entropy of mean, and the average of the standard deviation of all Gaussians from the latent space of the $\beta$VAE model. (d) shows the value range, CI, and entropy of the AE.}
 \label{fig:empty_latent}
\end{figure}
\setlength{\belowcaptionskip}{0pt}

Switching to the AE for comparison (R4.2), we could not find any deprecated dimensions. As shown in \autoref{fig:empty_latent}d, all dimensions have a wide value range and large entropy (of the latent values).

\setlength{\intextsep}{10pt}%
\setlength{\columnsep}{15pt}%
\begin{wrapfigure}{r}{0.4\linewidth}
 \centering
 \includegraphics[width=0.4\columnwidth]{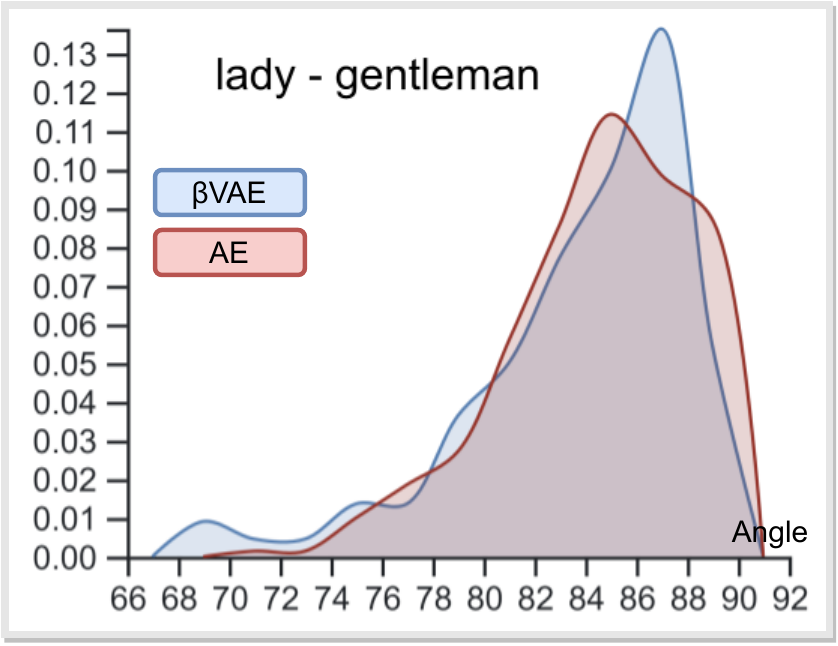}
 \caption{Normalized angle distribution:  $\beta$VAE (blue) and AE (red).}
 \label{fig:angle_distribution}
\end{wrapfigure}

\paragraph{\textbf{Semantics' Encoding-Level.}}
Next, the exploration focuses on the encoding-level of a user-proposed semantics (R2.2, R2.3). As gender is a very common testbed for semantics explorations, we probe its encoding-level across the latent dimensions. We use the word pair \texttt{lady}-\texttt{gentleman} to represent this semantics. Note that, \texttt{woman}-\texttt{man} is a more commonly used pair for gender. We did not choose it because our later interpretation on that pair (i.e., the \textit{Word Cloud View}) involves words with gender discrimination.

After entering the two words in the \textit{Dimension Exploration view} (the two text boxes, \autoref{fig:system}b), their latent vectors are shown as two curves in the zoomable PCP. The semantics perturbation on each latent dimension is computed automatically and the angular glyph above each PCP axis is updated. To obtain an overview of the angles, we click the ``Angle Distribution'' button in \autoref{fig:system}-b4, and the normalized angle distributions for $\beta$VAE (110 angles for the 110 useful dimensions) and AE (350 angles) are shown in \autoref{fig:angle_distribution}. It is obvious that the latent dimensions of $\beta$VAE either have a smaller or a larger angle (blue area) than that of the AE (red area). The smaller angle dimensions (the lower-left corner) are dimensions encoding more gender semantics, and the larger angle ones (the top-right corner) are dimensions encoding less gender semantics. 
This indicates that $\beta$VAE regularizes the semantics' encoding in latent dimensions by condensing the semantics into fewer dimensions (the lower-left ones), and freeing some of the rest dimensions, such that they become more orthogonal to the semantics (the top-right ones).

\paragraph{\textbf{Generalizability.}} Using our system, we also explored the latent space of AE and $\beta$VAE for other languages (R4.3), e.g., Spanish, Italian, and German. 
In all languages, our compressed embeddings from $\beta$VAE achieve similar quality (i.e., similar semantic similarity and analogy score) with the 350D embeddings from AE, but use much fewer dimensions, validating that our $\beta$VAE effectively removes redundant information encoded in AE.

We have also explored the angle distributions using different semantics of the same language, e.g., \texttt{yes}-\texttt{no} and \texttt{beautiful}-\texttt{ugly}, or the same semantics in different languages, e.g., \texttt{señora}-\texttt{caballero} and \texttt{signora}-\texttt{signore} (the Spanish and Italian counterparts for \texttt{lady}-\texttt{gentleman}). A common trend across the explorations is that the latent dimensions with smaller angles usually come from $\beta$VAE (similar to the lower-left corner of \autoref{fig:angle_distribution}), indicating those dimensions are more semantically salient. In some cases, the AE may have more dimensions with around 90 degrees (different from the top-right corner of \autoref{fig:angle_distribution}). This is also reasonable, as some latent dimensions of AE indeed encode little information of the probed semantics due to their chaotic encoding.

\subsection{Semantic Analysis}
While exploring the latent space, several dimensions of interest were identified. This section further investigates/compares them (R3).

\paragraph{\textbf{Deprecated Dimension.}}
The first thing that we were eager to verify is whether a deprecated dimension is really deprecated. By perturbing the latent vector from a deprecated dimension, we visualize the reconstructions in the \textit{Embedding Projection View}. As shown in \autoref{fig:deprecated_dimension}a, the asterisk shows the projection of the reconstructed embedding for \texttt{lady}. 
We perturbed the deprecated dimension to generate 700 samples, but all the reconstructions for these perturbations are overlaid on the dark blue point. This means no matter how we perturb the latent value on the deprecated dimension, the reconstruction will remain the same, i.e., the deprecated dimension is indeed useless. In comparison, \autoref{fig:system}-c2 shows the reconstructions from the perturbation of a useful dimension. The reconstructions are distributed linearly, and we use the color from dark blue to black to indicate the reconstructions' distance to the asterisk. Similar observations are also found for \texttt{gentleman} (\autoref{fig:system}-c1, \ref{fig:deprecated_dimension}a).

\setlength{\belowcaptionskip}{0pt}
\begin{figure}[tbh]
 \centering
 \includegraphics[width=\columnwidth]{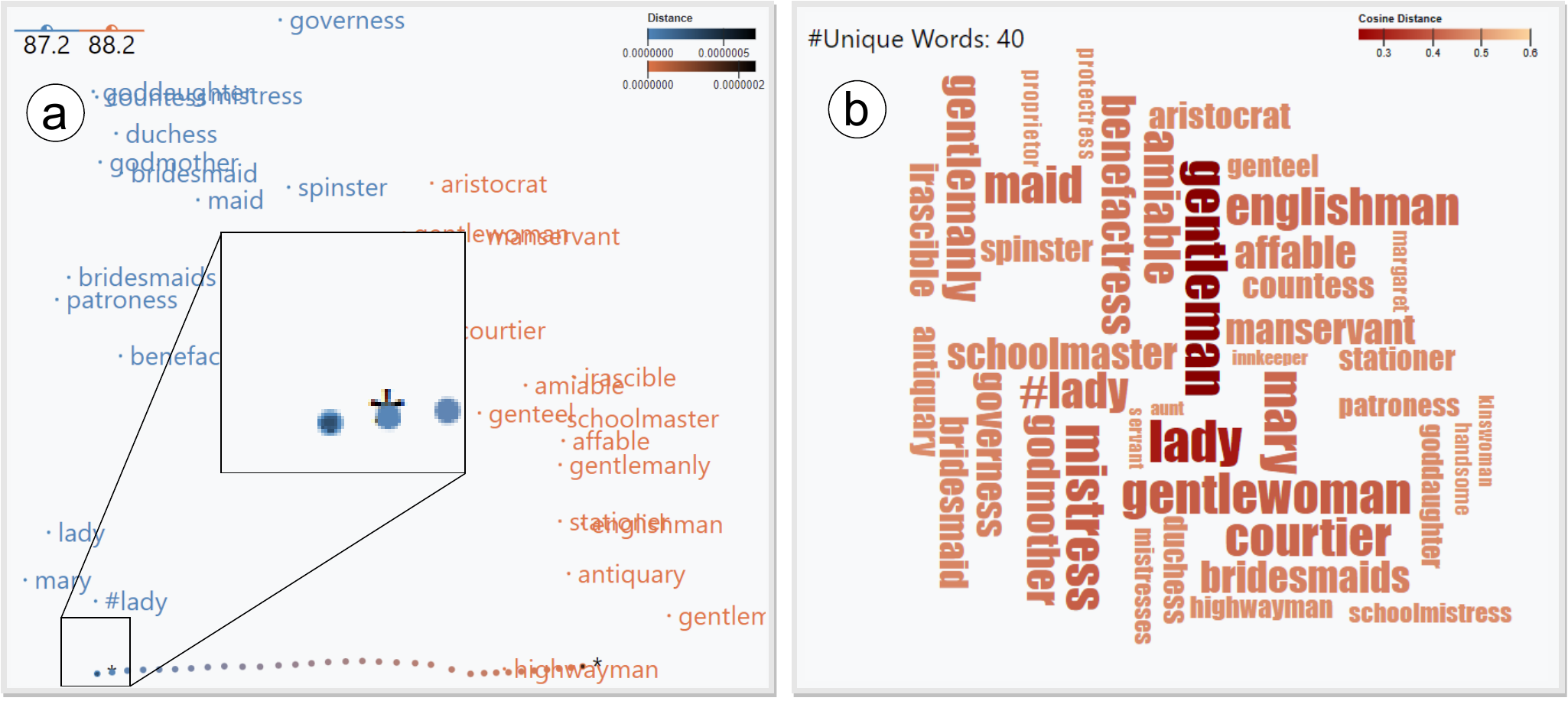}
 \caption{The \textit{Embedding Projection View} (a) and \textit{Word Cloud View} (b) of the perturbed samples for a deprecated dimension.}
 \label{fig:deprecated_dimension}
\end{figure}
\setlength{\belowcaptionskip}{0pt}

\autoref{fig:system}d and \autoref{fig:deprecated_dimension}b show the \textit{Word Cloud View} when brushing the full range of the useful and deprecated dimension. They have 70 and 40 unique words respectively, indicating the useful dimension encodes more variance of semantics. Note that the word cloud is generated by finding nearest neighbors of the perturbed samples (explained in \autoref{fig:cloud}), so this view for the deprecated dimension still has multiple words. However, it is obvious that the words in \autoref{fig:deprecated_dimension}b are smaller and in a lighter color (compared to \autoref{fig:system}d), indicating the space is sparser and the word frequency is lower. For both cases, especially \autoref{fig:system}d, we found meaningful words with larger size and darker color, helping to textualize the surrounding space of the studied semantics, e.g., \texttt{gentlewoman} and \texttt{englishman}.

\paragraph{\textbf{Small and Large Angle Dimensions.}} Next, focusing on useful dimensions, we compare dimensions with a small and large angle to identify their difference regarding the studied semantics. \autoref{fig:system}c and \ref{fig:system}d show the visualization of a latent dimension with a small angle, whereas  \autoref{fig:large_small_angle} shows the same visualization but for the dimension with a large angle. Despite the distortion in the \textit{Embedding Projection View}, the angles in the 2D projection (\autoref{fig:system}c, \autoref{fig:large_small_angle}a) correspond well with the angles in the original HD space (as our perturbation is localized around a word along a single dimension). The top-left angular glyphs also precisely show the angles. The experts appreciated the intuition this view brought.

The two corresponding \textit{Word Cloud Views} reveal more semantics. For example, compared to the small angle dimension (\autoref{fig:system}d), the large angle dimension (\autoref{fig:large_small_angle}b) encodes a little more unique words, and we highlight some of them with the blue dotted squares, e.g., \texttt{levett}, \texttt{pepys}, \texttt{elizabeth}, and \texttt{lancelot}. These words are common noble last names.
Since \texttt{lady}-\texttt{gentleman} are usually used to refer to people of nobility, we speculate the chosen large angle dimension may encode semantics to some extent related to well-known names, which is orthogonal to the gender semantics (explaining the dimension's nearly 90\textdegree~angle with \texttt{lady}-\texttt{gentleman}).

\setlength{\belowcaptionskip}{-12pt}
\begin{figure}[tb]
 \centering
 \includegraphics[width=\columnwidth]{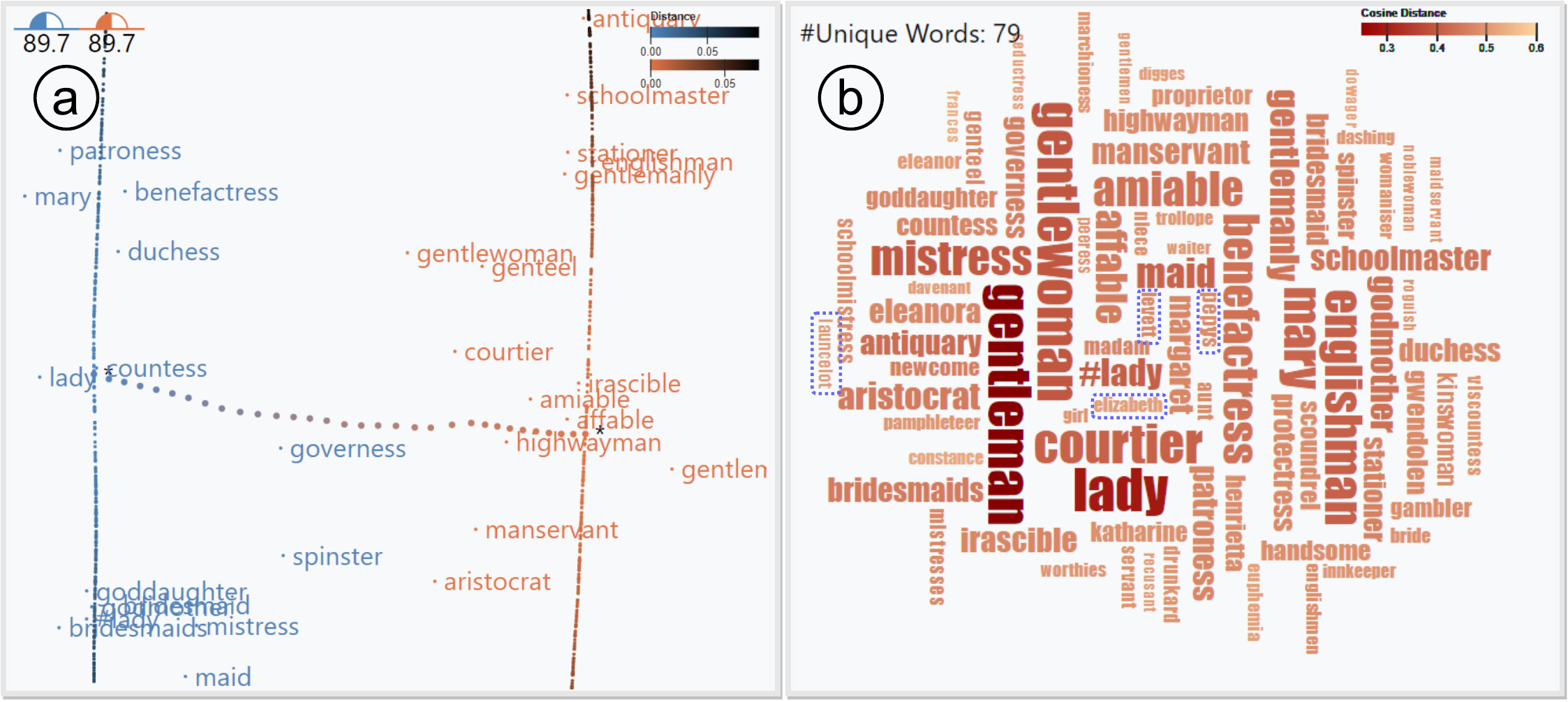}
 \caption{ (a) The \textit{Embedding Projection View} (a) and \textit{Word Cloud View} (b) of a large angle dimension. The four words highlighted in (b) are  \texttt{levett}, \texttt{pepys}, \texttt{elizabeth}, and \texttt{lancelot}. }
 \label{fig:large_small_angle}
\end{figure}
\setlength{\belowcaptionskip}{0pt}

\begin{figure}[tbh]
 \centering
 \includegraphics[width=\columnwidth]{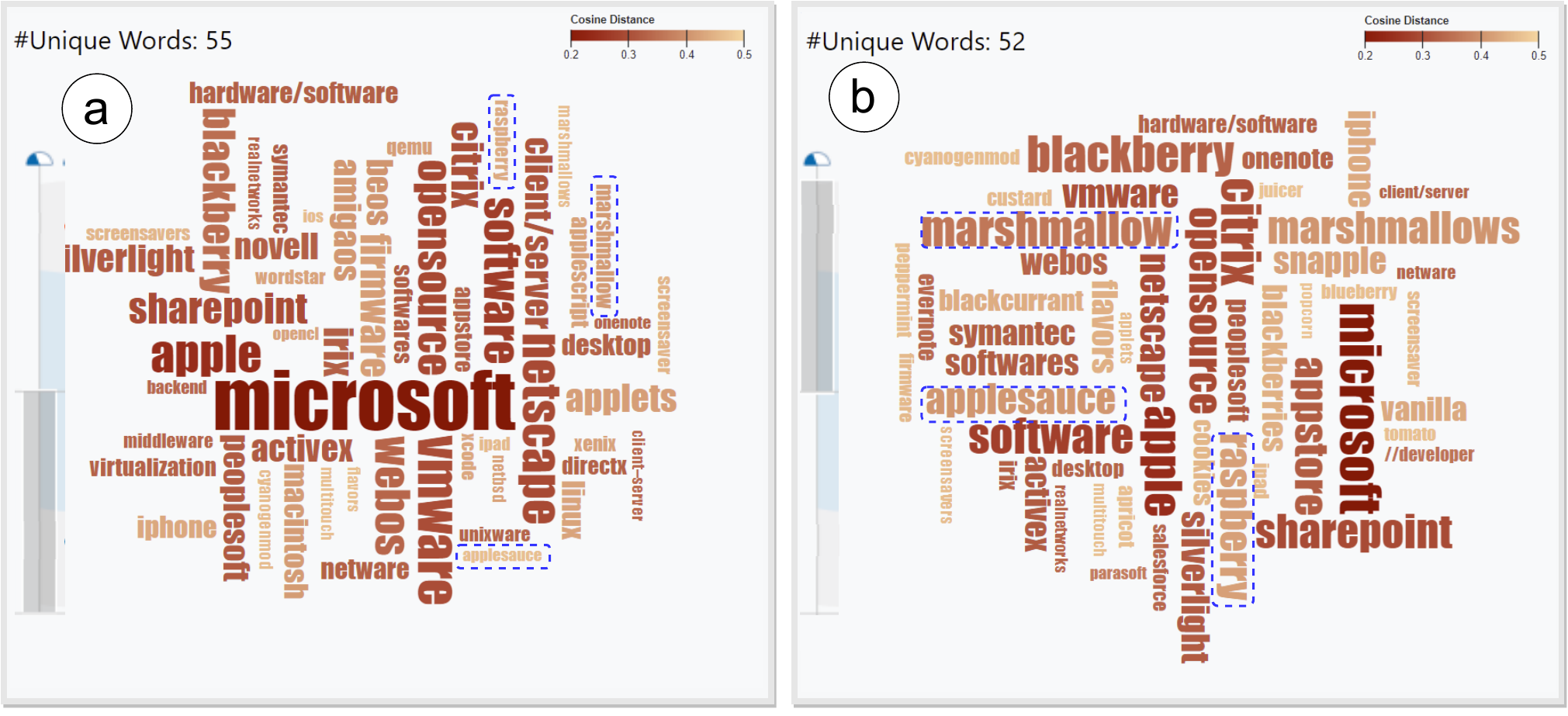}
 \caption{Brushing different ranges of a latent dimension to explore the semantics transition with the \textit{Word Cloud View}.
 }
 \label{fig:word_cloud_segment}
\end{figure}

\paragraph{\textbf{Semantics Transitions along One Dimension.}}
The transition analysis of \texttt{lady}-\texttt{gentleman} is rather subtle, making it less appropriate to demonstrate the use of our system. 
Instead, we use another word pair \texttt{apple}-\texttt{microsoft} to show a clear semantics transition along a latent dimension and include the analysis of  \texttt{lady}-\texttt{gentleman} in our Appendix. The semantic ambiguity of \texttt{apple}, which can refer to a kind of fruit or a technology company, makes it interesting to explore. 
We demonstrate a possible dimension involving this semantics transition through word clouds generated by probing different value ranges of the dimension: $min$-$median$ (\autoref{fig:word_cloud_segment}a), and $median$-$max$ (\autoref{fig:word_cloud_segment}b). The gray and blue bands on the zoomable PCP help to brush these ranges. Comparing the two word clouds, \texttt{raspberry}, \texttt{applesauce}, and \texttt{marshmallow}, which are related to the fruit meaning of \texttt{apple}, are of higher-frequency in \autoref{fig:word_cloud_segment}b, while \autoref{fig:word_cloud_segment}a mostly shows words related to technology.

\setlength{\intextsep}{10pt}%
\setlength{\columnsep}{18pt}%
\begin{wrapfigure}{r}{0.48\linewidth}
 \centering
 \includegraphics[width=0.46\columnwidth]{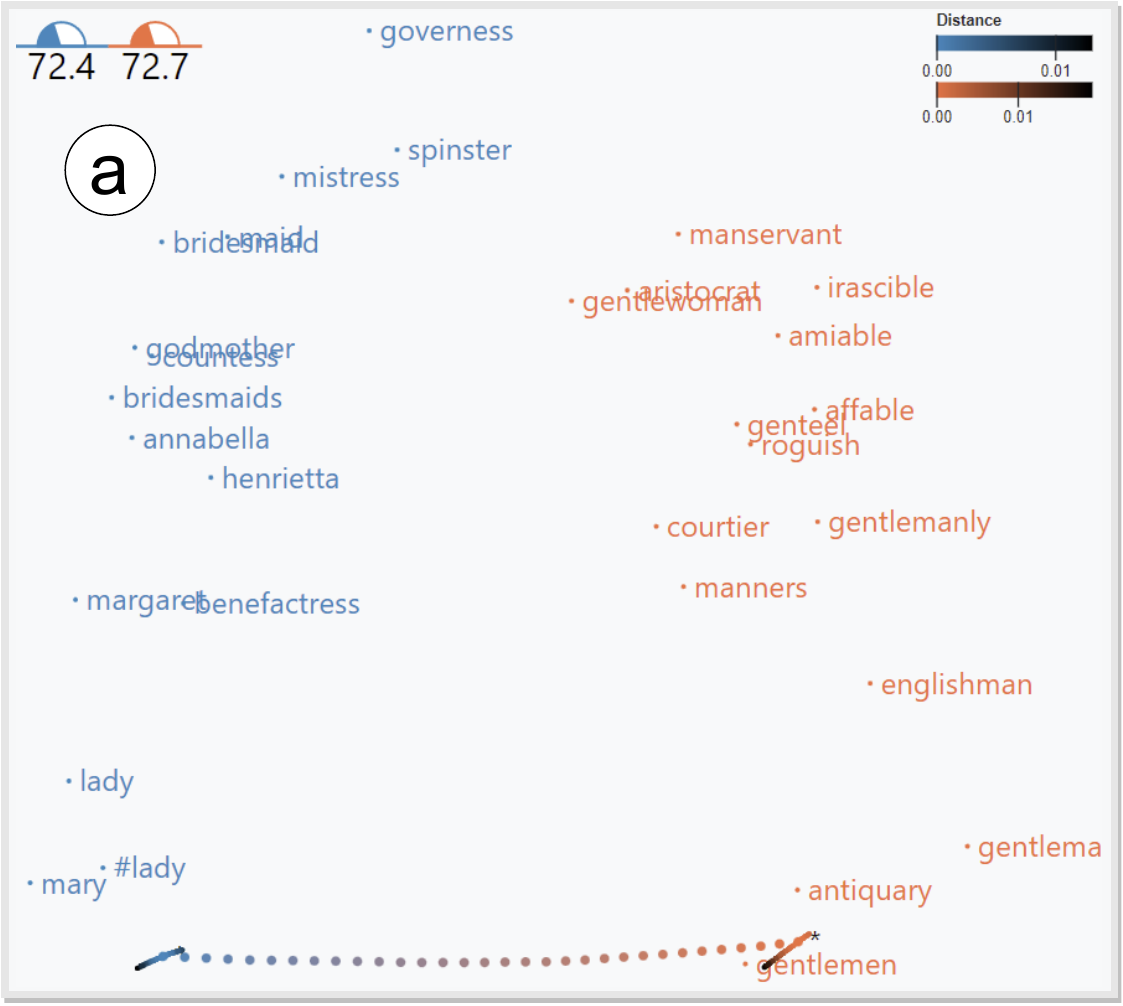}
 \caption{Latent dimension (with a small angle) from AE.}
 \label{fig:ae_projection_word_cloud}
\end{wrapfigure}
\paragraph{\textbf{Comparison with AE.}} To compare with AE, we select AE's latent dimension with the smallest angle as well. From the \textit{Embeddding Projection View} (\autoref{fig:ae_projection_word_cloud}), the perturbation of this AE dimension usually has smaller extension (compared to \autoref{fig:system}-c1, c2). The corresponding word cloud also has fewer unique words (47 compared to 70 in \autoref{fig:system}d). Similar findings have also been observed when exploring other AE dimensions. They indicate the AE encodes embeddings into a higher dimensional but much sparser/diluted space.

\subsection{Domain Experts' Feedback}
\label{sect:feedback}
Following the \textit{guided exploration + think-aloud discussion} protocol, we went through the above cases with the 4 experts introduced in the Design Requirement Analysis ($E_1{\sim}E_4$). The case studies were carried out in two separate sessions (one hour each), one focuses on explaining our solutions, the other focuses on exploring the cases and open-discussions.

In general, all experts provide positive feedback on our system in presenting the regularization process, exploring the HD latent space, and interpreting individual dimensions' semantics. $E_1$ felt the interpolation between words' latent representations is interesting and he agreed the \textit{Embedding Projection View} is intuitive for the understanding of our semantics probing process. He also commented that dimensions with a `${<}75$' degree angle are usually meaningful and glad to see that the dimension with the smallest angle from $\beta$VAE is smaller than that from AE. $E_1{\sim}E_3$ were surprised to see the clear distinction between useful and deprecated latent dimensions from the \textit{Dimension Exploration View} and appreciated our quantitative evaluations to validate the on-par quality of the regularized embeddings. We also discussed the linear distribution of the perturbed samples with them (e.g., \autoref{fig:system}-c1). As our perturbations are around a given word along a single dimension, the locality contributes significantly to the linear layout of the projected reconstructions. Inspired by the insights from our work, $E_4$ proposed the idea of alternatively training and regularizing word embeddings, in which, humans can steer/optimize the embeddings for specific downstream tasks. 

Our discussions also include some limitations and desired features of the system. First, when comparing two \textit{Word Cloud Views}, the unique words from them could be small and hard to identify. Hiding the common words flexibly between two word clouds may help here.
Second, two experts brought up the capability to explore multiple latent dimensions concurrently, e.g., perturbing all dimensions with `${<}75$' degree angles as they all have high encoding-level of the studied semantics. This capability would require more sophisticated perturbation mechanisms and interaction solutions, which we plan to explore more to extend our system. 

\section{Quantitative Evaluations}
\label{sect:evaluation}

This section presents quantitative evaluations to rigorously validate the efficacy of our approach in embedding regularization.

\paragraph{\textbf{Monolingual Evaluation}}
For monolingual evaluation, we train an AE and a $\beta$VAE on a language's embeddings (300D from FastText\cite{bojanowski2017enriching}) and use the corresponding latent representations as new embeddings to compare the encoding quality of the AE and $\beta$VAE.
\clrb{Since the focus of this evaluation is to compare the embedding quality after transformation, we only show the raw embedding quality (FastText) as a quality upper bound.}
We use five languages in this experiment, English, German, Italian, Spanish, and Persian. 
Both AE and $\beta$VAE are trained with 350 latent dimensions. 

In \autoref{tab:single_lang}, the semantic similarities of the original FastText embeddings, the latent representations from AE and $\beta$VAE are shown as ``FastText", ``AE350" and ``$\beta$VAE" respectively. The subscript in ``$\beta$VAE" is the number of \textit{useful} dimensions of the corresponding $\beta$VAE, which is about 1/3 of the original 350 dimensions (around 120). The similarity computation in $\beta$VAE uses these useful dimensions only. Based on this, we further train another AE with 120 latent dimensions, the result is shown in the row of ``AE120".
Comparing ``AE350" and ``$\beta$VAE", ``$\beta$VAE" achieves comparable performance, but significantly reduces the number of latent dimensions (i.e., the deprecated dimensions are indeed useless). Comparing ``$\beta$VAE" and ``AE120", which share a similar number of latent dimensions, ``$\beta$VAE" better preserves the embeddings' quality.

\begin{table}[tb]
  \caption{Semantics evaluation using the SemEval 2017~\cite{camachocollados-EtAl:2017:SemEval}.}
  \label{tab:single_lang}
  \fontsize{8}{8}\selectfont
	\centering
  \begin{tabularx}{\columnwidth}{cccccc}
  \toprule
       & English    & German    & Italian    & Spanish    & Persian    \\
  \midrule 
       FastText  & 0.729 & 0.735 &0.737 & 0.749  & 0.678 \\
    \midrule
       AE350  & 0.698 & \textbf{0.722} & 0.726 & \textbf{0.749}  & 0.668\\
       $\beta$VAE & $\textbf{0.708}_{110}$ & $0.713_{123}$ &$\textbf{0.737}_{110}$ & $0.745_{110}$  & $\textbf{0.693}_{103}$ \\
       AE120  & 0.691 & 0.711 & 0.711 & 0.743  & 0.660 \\
  \bottomrule
  \end{tabularx}
\end{table}

Focusing on English, we further profile the effect of $\beta$ in $\beta$VAE. Specifically, we trained multiple $\beta$VAE with 350 latent dimensions but different $\beta$ values. After the models convergence, we found a clear separation of the useful and deprecated dimensions in all models.
\autoref{tab:beta} (left) reports the number of useful dimensions, the similarity and analogy scores (computed using the useful dimensions only). 
The general trend is that a larger $\beta$ enforces a stronger constraint on the latent space, forcing more dimensions to become deprecated and concentrating information into fewer dimensions. However, this also comes with a cost of the embedding quality.
In our studies, we apply the random search algorithm to determine the hyperparameter $\beta$ for neural network training, which is widely used in neural network designs. $\beta{=}10^{-5}$ is found to be the sweet point for all our tested languages.

\begin{table}\fontsize{7.5}{8}\selectfont
\centering
\captionsetup[subfloat]{labelformat=empty}
\caption{(Left) The effect of different $\beta$ values. (Right) Average BLI accuracy (\%).}
\label{tab:beta}
\subfloat[]{%
\begin{tabular}{p{0.6cm}p{0.5cm}p{0.5cm}p{0.5cm}p{0.5cm}}
  \toprule
    & 1e-5 & 3e-5  & 5e-5  & 7e-5  \\
  \midrule 
    Dim.   & 110   & 76    & 60    & 52     \\
    SemEval     & 0.708 & 0.685 & 0.669 & 0.658  \\
    Analogy     & 0.636 & 0.434 & 0.446 & 0.423  \\
  \bottomrule
\end{tabular}}%
\quad 
\subfloat[]{%
\begin{tabular}{p{0.4cm}p{0.3cm}p{0.3cm}p{0.5cm}}
\toprule
  & 1NN & 5NN & 10NN  \\
  \midrule 
AE350 & 55.9 & 70.6 & 75.0 \\
$\beta$VAE & 54.8 & 69.0 & 73.6  \\
AE120 & 49.9 & 64.2 & 69.3  \\
\bottomrule
\end{tabular}}
\end{table}

\paragraph{\textbf{Cross-lingual Evaluation}}
We further evaluate the quality of the latent word embeddings on a downstream task of word translation (BLI). Specifically, we translate between English and either other languages. For each language pair, i.e., English and * (* for one of the eight languages), we follow the same settings with LNMap~\cite{mohiuddin2020lnmap} to (1) encode them into a latent space and (2) align them through two mapper functions and fine-tuning of the encoder networks. For step (1), we use ``AE350'', ``$\beta$VAE'', and ``AE120'' to encode the two languages' embeddings (note that AE350 is the original LNMap).

After the alignment, we calculate translation accuracy when considering 1, 5, and 10 nearest neighbors (NN) for both sources to target and target to source mapping. The details of these accuracy numbers can be found in our Appendix. \autoref{tab:beta} (right) shows the average accuracy across all 8 language pairs.
\clrb{In short, ``$\beta$VAE'' is slightly worse than ``AE350'', but significantly better than ``AE120''.}

\section{Discussion, Limitations, and Future Work}
\clrb{
Although the quantitative results show the word embeddings generated with $\beta$VAE are slightly worse than the baseline method ``AE350'', a smaller number of dimensions and less entangled semantics make it easier to analyze the resulting word embeddings. Meanwhile, with a similar number of embedding dimensions, our method consistently outperforms ``AE120'' in embedding quality. In short, our method has a better embedding compression and regularization quality than the traditional AE methods. 
}

\clrb{
Our approach stems from the phenomenon of dimension deprecation in the training of $\beta$VAE and we have observed this phenomenon across all tested language embeddings. 
Although we cannot guarantee a clear separation between useful and deprecated dimensions in all types of embeddings, an entropy threshold can always be used to filter out the dimensions with more information, and other parts of our approach would remain applicable.}

\clrb{
There are several \textit{\textbf{limitations}} in our current work. \textit{First}, similar to existing latent space interpretation works~\cite{liu2019latent}, our work also relies on the linear assumption of semantics in the HD latent space and thus uses the first principle component's direction from PCA to regress the perturbed samples. For our studied problem, as our perturbation is conducted locally, this assumption works well and we generate meaningful interpretations. However, it is still worth more theoretical studies on this topic, which our collaborated domain experts are actively working on. \textit{Second}, our system lacks informative hints to guide users in exploring a large number of latent dimensions. Although we can easily distinguish the useful dimensions from the deprecated ones, the number of useful dimensions could still be large and most of them may have similar angles to the studied semantics.
Consequently, which useful dimension deserves more attention from the experts is still an open question. We believe some metrics should be proposed here to profile individual latent dimensions and provide more guidance during explorations, and plan to work in this direction to improve our system. \textit{Third}, for the qualitative evaluation part, we rely on the case studies with experienced domain experts. How users with different backgrounds and expertise would like our system has not been rigorously evaluated. Therefore, one part of our future work is to conduct statistical user studies with users of different levels of NLP knowledge.
}

\clrb{
Our future works also include the exploration and comparison of more VAE architectures and other ways to regularize the latent space. As discussed with the domain experts, humans can introduce certain supervisions to guide the training of VAE or $\beta$VAE to regularize the encoding of certain semantics of interest, where interactive visual exploration will play an important role. Also, as mentioned in \hyperref[sect:feedback]{Domain Experts' Feedback}, we plan to study how to perturb multiple latent dimensions concurrently to probe and interpret their collective behaviors and develop a friendly visual interface for user interactions.
}

\section{Conclusion}
\clrb{
In this paper, we propose using $\beta$VAE to compress and regularize HD word embeddings, and probing the encoding-level of different semantics encoded in each latent dimension through interactive perturbations. With our visual analytics system, we can easily disclose how $\beta$VAE regularizes the HD latent space of word embeddings by concentrating information into fewer latent dimensions and deprecating superfluous ones. Also, our system supports users to explore and compare user-proposed semantics encoded in each latent dimension, and investigate the semantics transition along a dimension.
Through both quantitative and qualitative evaluations, we rigorously verified the superior performance of our $\beta$VAE-based embedding compression and regularization solution and proved that the individual $\beta$VAE latent dimensions are more semantically salient.
}




\bibliographystyle{SageV}
\bibliography{main.bib}

\end{document}